\newif\ifmixed
\mixedtrue
\documentclass[sigconf,nonacm]{acmart}

\usepackage{mathtools}
\usepackage{booktabs}
\usepackage{array}
\usepackage{microtype}
\definecolor{KeyBlue}{HTML}{EAF3F8}
\definecolor{KeyLine}{HTML}{2B6F92}
\newtheorem{assumption}{Assumption}
\newtheorem{proposition}{Proposition}
\newtheorem{corollary}{Corollary}
\newcommand{\R}{\mathbb{R}}
\newcommand{\one}{\mathbf{1}}
\newcolumntype{L}[1]{>{\raggedright\arraybackslash}p{#1}}

\title{Penalize the Common Trade: A Collision-Aware Layer for Learned Portfolios}

\setcopyright{none}
\makeatletter\global\@ACM@balancefalse\makeatother
\acmConference{}{}{}
\acmBooktitle{}
\acmDOI{}
\acmISBN{}

\begin{document}

\author{Eric Benhamou}
\affiliation{
  \institution{Ai For Alpha}
  \city{Neuilly-sur-Seine}
  \country{France}
}
\affiliation{
  \institution{LAMSADE, Universit\'e Paris Dauphine--PSL}
  \city{Paris}
  \country{France}
}
\email{eric.benhamou@aiforalpha.com}

\author{Jean-Jacques Ohana}
\affiliation{
  \institution{Ai For Alpha}
  \city{Neuilly-sur-Seine}
  \country{France}
}
\email{jean-jacques.ohana@aiforalpha.com}

\author{Beatrice Guez}
\affiliation{
  \institution{Ai For Alpha}
  \city{Neuilly-sur-Seine}
  \country{France}
}
\email{beatrice.guez@aiforalpha.com}

\author{Ethan Setrouk}
\affiliation{
  \institution{Ai For Alpha}
  \city{Neuilly-sur-Seine}
  \country{France}
}
\email{ethan.setrouk@aiforalpha.com}

\author{Chamyl Saadi}
\affiliation{
  \institution{Ai For Alpha}
  \city{Neuilly-sur-Seine}
  \country{France}
}
\email{chamyl.saadi@aiforalpha.com}

\author{Thomas Jacquot}
\affiliation{
  \institution{Ai For Alpha}
  \city{Neuilly-sur-Seine}
  \country{France}
}
\email{thomas.jacquot@aiforalpha.com}
\renewcommand{\shortauthors}{Benhamou et al.}

\begin{abstract}
Different learning systems can make nearly the same trade when they share data, objectives, or representations. The market then sees decision concentration even if the models look diverse. We call exposure to this shared direction a \emph{signal collision}. A linear-impact game shows the basic externality: each allocator chooses too much common exposure because it does not bear the full price effect imposed on others. An overlap-dependent toll recovers the social choice, and a feedback model gives the exact gain at which return-responsive allocation becomes unstable. We also state the restrictive spectral conditions under which a leading eigenportfolio can reveal the collision direction.

These results motivate a simple post-model layer: subtract a penalty for loading on the recent common return mode before forming the portfolio. In a frozen panel of ten liquid cross-asset ETFs, 760 completed weeks from January 2012 through July 2026, and exact self-financing costs of 10 basis points per dollar traded, this layer changes maximum drawdown from 23.6\% to 14.1\%, volatility from 12.6\% to 10.7\%, and excess-return Sharpe from 0.664 to 0.667. A volatility-matched momentum diagnostic still draws down 19.9\%. Paired stationary-bootstrap uncertainty is wide: the 95\% interval for drawdown improvement is $[-3.3,18.4]$ percentage points and includes zero. The result establishes a reproducible risk-control association and specifies the holdings-and-flow validation needed to attribute the common direction to AI crowding.

\par\smallskip\noindent\textbf{AI-use disclosure.} Generative AI tools, including OpenAI Codex, assisted manuscript preparation. For this preprint version, Codex assisted with LaTeX editing, author and venue metadata, compilation, and layout checks.
\end{abstract}

\maketitle
\hypersetup{pdfauthor={Eric Benhamou; Jean-Jacques Ohana; Beatrice Guez; Ethan Setrouk; Chamyl Saadi; Thomas Jacquot}}
\begin{center}\small Author preprint.\end{center}

\section{The Market Sees Trades, Not Model Names}

Imagine two funds built by different teams. One uses a transformer, the other an ensemble, yet both read similar prices and filings and optimize similar losses. If both systems buy the same assets after good news and sell them after losses, their source code is diverse but their market action is not. The important object is therefore the \emph{common trade}, not similarity between model architectures.

This observation matters when trades move prices. Ordinary factor exposure describes a shared response to an external return process. A signal collision can instead change that process: aligned purchases consume liquidity, and aligned exits can deepen losses. Common holdings and forced sales are established sources of fragility \cite{GreenwoodThesmar2011,Greenwood2015,Caccioli2014}; funding constraints and the August 2007 quant episode show how quickly common strategies can unwind \cite{ShleiferVishny1997,BrunnermeierPedersen2009,KhandaniLo2011}. Learned allocators add a plausible coordination channel because data, representations, and objectives are often reused.

\begin{center}
\fcolorbox{KeyLine}{KeyBlue}{\parbox{\dimexpr\columnwidth-2\fboxsep-2\fboxrule\relax}{\small\textbf{Key message.} Model diversity does not guarantee trade diversity. A transparent post-model layer prices the common trade, reduces shared-direction loading, and yields a direct validation agenda based on holdings and flows.}}
\end{center}

We ask one focused question: \emph{if a learned portfolio proposal is crowded, what correction is economically justified, and what can public returns tell us about it?} Four answers organize the paper.
\begin{enumerate}
    \item \emph{Why correct the proposal?} A linear-impact game separates private and social common exposure and derives the overlap toll that closes the gap.
    \item \emph{When can crowding destabilize learning?} A closed feedback recursion gives the exact loop-gain threshold and the attenuation from a common-action penalty.
    \item \emph{What can prices identify?} A separated-spike result gives conditions under which the leading covariance eigenvector tracks the latent direction and identifies the additional holdings evidence required to interpret it as AI crowding.
    \item \emph{Does the layer change an observed path?} A frozen public-data experiment audits one rank wrapper with self-financing costs, placebos, asset omissions, a full penalty grid, and paired bootstrap uncertainty.
\end{enumerate}

The theory and experiment answer complementary questions. Conditional on a shared direction, the theory identifies the externality and its correction. The experiment measures how avoiding a recent common return mode changes one momentum strategy's risk. Holdings and order flow form the next validation layer that joins the reduced-form direction to the structural mechanism.

\section{What Is Inherited, and What Is New}

The paper connects three mature literatures. Greenwood and Thesmar \cite{GreenwoodThesmar2011} link concentrated ownership and correlated liquidity shocks to price fragility; Greenwood, Landier, and Thesmar \cite{Greenwood2015} quantify vulnerability through common exposures; and Caccioli et al. \cite{Caccioli2014} model contagion through overlapping portfolios. Cont and Wagalath \cite{ContWagalath2016} recover fire-sale pressure from returns. We express this externality as a post-model correction: the base learner proposes a score, and the collision layer discounts the direction most likely to be shared.

The feedback mechanism follows the logic of liquidity spirals, in which losses induce trades that cause further losses \cite{BrunnermeierPedersen2009}. Our contribution is an interpretable stability threshold for a stated return-responsive rule and a closed-form attenuation factor for the decision layer.

The empirical base is momentum, whose evidence and implementation risks are well known \cite{JegadeeshTitman1993,Moskowitz2012}. Principal components summarize covariance, while Davis--Kahan theory controls eigenvector rotation only when a population direction is sufficiently separated \cite{DavisKahan1970,Yu2015}. Our innovation is the bridge from an externality model to an auditable, model-agnostic decision layer. Explicit spectral conditions and falsification tests keep the return principal component in its correct role as a \emph{collision proxy}; holdings overlap supplies the structural validation target.

\section{A Small Game Makes the Externality Visible}

We begin with the smallest model that separates a good signal from an overcrowded trade. There are $N\ge2$ mean-variance allocators. Allocator $i$ observes the same signed signal $s$ and chooses scalar exposure $x_i\in\R$ to its normalized direction. The market sees their total $Q=\sum_{j=1}^N x_j$. Conditional return is
\begin{equation}
 r=\theta s-\lambda Q+\varepsilon,\qquad
 \mathbb E[\varepsilon\mid s]=0,\quad
 \operatorname{Var}(\varepsilon\mid s)=\sigma^2,
 \label{eq:return}
\end{equation}
Here $\theta s$ is the pre-impact expected return, $\lambda>0$ is adverse price impact, and $\varepsilon$ is residual news. A second coefficient $\eta\ge0$ summarizes expected liquidation loss proportional to common exposure; write $L=\lambda+\eta$. With risk aversion $\gamma>0$, allocator $i$ maximizes
\begin{equation}
 u_i(x_i,Q_{-i})=x_i\theta s-Lx_i(x_i+Q_{-i})
 -\frac{\gamma\sigma^2}{2}x_i^2,
 \label{eq:utility}
\end{equation}
where $Q_{-i}=\sum_{j\ne i}x_j$ is fixed when allocator $i$ deviates. The term $-Lx_i(x_i+Q_{-i})$ is the key: the allocator feels the cost on its own position, but not the full loss its order causes for everyone else. The quadratic form is a local benchmark; nonlinear impact and inventory constraints lie outside the result.

\begin{proposition}[Private exposure, social exposure, and toll]
\label{prop:game}
Game \eqref{eq:utility} has a unique Nash equilibrium. It is symmetric, with
\begin{equation}
 x^{\mathrm{NE}}=\frac{\theta s}{\gamma\sigma^2+L(N+1)}.
 \label{eq:ne}
\end{equation}
The unique maximizer of aggregate utility is symmetric, with
\begin{equation}
 x^{\mathrm{SO}}=\frac{\theta s}{\gamma\sigma^2+2LN}.
 \label{eq:so}
\end{equation}
For $s\ne0$, $|x^{\mathrm{NE}}|>|x^{\mathrm{SO}}|$. Subtracting the toll
$Lx_iQ_{-i}$ from each private utility implements $x^{\mathrm{SO}}$ as the unique Nash equilibrium.
\end{proposition}

\begin{proof}
The first-order condition is
$\theta s-(2L+\gamma\sigma^2)x_i-LQ_{-i}=0$.
The associated linear system has positive eigenvalues
$\gamma\sigma^2+L$ in directions orthogonal to $\one$ and
$\gamma\sigma^2+L(N+1)$ in the common direction; hence its solution is unique and gives \eqref{eq:ne}. Aggregate utility is
$\theta sQ-LQ^2-(\gamma\sigma^2/2)\sum_i x_i^2$, a strictly concave function. Its first-order conditions give \eqref{eq:so}. Since $2N>N+1$, the social denominator is larger. The toll contributes $-LQ_{-i}$ to the private first-order condition, changing its symmetric denominator to
$\gamma\sigma^2+L(N+1)+L(N-1)=\gamma\sigma^2+2LN$.
\end{proof}

The denominators reveal the economics. A private allocator prices roughly one copy of the crowding it creates, whereas the planner prices the effect across all $N$ positions. The toll $Lx_iQ_{-i}$ charges only overlap with other allocators, so it vanishes when the trade is not shared. This is why a generic shrinkage penalty and a collision penalty are not the same object.

\begin{corollary}[Aggregate wedge and welfare loss]
\label{cor:welfare}
Write $a=\theta s$ and $h=\gamma\sigma^2$. Aggregate welfare at a symmetric profile $x\one$ is $W(x)=Na x-LN^2x^2-(hN/2)x^2$. Hence
\begin{align}
 W(x^{\mathrm{SO}})-W(x^{\mathrm{NE}})
 &=\frac{N(h+2LN)}{2}(x^{\mathrm{NE}}-x^{\mathrm{SO}})^2,\label{eq:welfare-gap}\\
 x^{\mathrm{NE}}-x^{\mathrm{SO}}
 &=\frac{aL(N-1)}{[h+L(N+1)][h+2LN]}.
 \label{eq:exposure-gap}
\end{align}
For fixed $a,h$, and $L>0$, as $N\to\infty$,
$Nx^{\mathrm{NE}}\to a/L$, $Nx^{\mathrm{SO}}\to a/(2L)$, and the welfare loss in \eqref{eq:welfare-gap} tends to $a^2/(4L)$.
For $a>0$, the aggregate wedge
$\Delta Q=N(x^{\mathrm{NE}}-x^{\mathrm{SO}})$ increases with integer $N\ge2$. As a function of $L$, however, it is hump-shaped and is maximized at
$L^*=h/\sqrt{2N(N+1)}$ when $h>0$.
\end{corollary}

\begin{proof}
Completing the square in the strictly concave quadratic $W$ gives \eqref{eq:welfare-gap}. Subtracting \eqref{eq:so} from \eqref{eq:ne} gives \eqref{eq:exposure-gap}; multiplying by $N$ gives $\Delta Q$. With $r=h/L$, the derivative in continuous $N$ of $N(N-1)/[(r+N+1)(r+2N)]$ has numerator $(3r+4)N^2+2r(r+1)N-r(r+1)>0$ for $N\ge1$. Differentiating $\Delta Q$ in $L$ has the sign of $h^2-2N(N+1)L^2$, which gives $L^*$; the limits follow after dividing each denominator by $N$.
\end{proof}

\begin{table}[t]
\caption{A numerical reading of the collision wedge. One normalized signal is shared by ten allocators ($a=h=1$, $L=0.1$, $N=10$).}
\label{tab:example}
\centering\small
\setlength{\tabcolsep}{3.5pt}
\renewcommand{\arraystretch}{1.06}
\begin{tabular}{L{.32\columnwidth}rL{.43\columnwidth}}
\toprule
Quantity & Value & Interpretation\\
\midrule
Private exposure $x^{\rm NE}$ & 0.476 & Each allocator prices only part of the overlap\\
Social exposure $x^{\rm SO}$ & 0.333 & The planner prices impact on all positions\\
Aggregate excess $N(x^{\rm NE}-x^{\rm SO})$ & 1.429 & Small individual gaps add up\\
Welfare loss $W(x^{\rm SO})-W(x^{\rm NE})$ & 0.306 & Loss in the model's utility units\\
\bottomrule
\end{tabular}
\end{table}

Table~\ref{tab:example} uses normalized units to expose the mechanism without market-specific calibration. Each allocator's excess is only 0.143, yet ten aligned choices create aggregate excess exposure of 1.429. The toll removes this gap by charging overlap while preserving the underlying signal.

Individual exposures vanish in a large market, but their aggregate externality need not. Equation~\eqref{eq:exposure-gap} also rules out a global monotonicity claim for the individual wedge: it is $O(N^{-1})$ as $N$ grows and tends to zero as $L\to\infty$. What persists is aggregate exposure, since $N(x^{\mathrm{NE}}-x^{\mathrm{SO}})\to a/(2L)$. The toll is cross-agent: it charges exposure only when others occupy the same direction.

\subsection{The same correction survives heterogeneity}

The implementation result extends to heterogeneous beliefs and risk costs. Let allocator $i$ have conditional expected return $\mu_i\in\R$ and quadratic cost $h_i>0$, and replace \eqref{eq:utility} by
\begin{equation}
 u_i=x_i\mu_i-Lx_i(x_i+Q_{-i})-\frac{h_i}{2}x_i^2.
 \label{eq:heterogeneous-utility}
\end{equation}
Set $H=\operatorname{diag}(h_1,\ldots,h_N)$ and $\mu=(\mu_1,\ldots,\mu_N)^\top$.

\begin{proposition}[Heterogeneous collision toll]
\label{prop:heterogeneous}
Game \eqref{eq:heterogeneous-utility} has the unique Nash exposure
\begin{equation}
 x^{\mathrm{NE}}=[H+LI+L\one\one^\top]^{-1}\mu.
 \label{eq:heterogeneous-ne}
\end{equation}
The unique aggregate-welfare maximizer is
\begin{equation}
 x^{\mathrm{SO}}=[H+2L\one\one^\top]^{-1}\mu.
 \label{eq:heterogeneous-so}
\end{equation}
Subtracting $Lx_iQ_{-i}$ from each private utility implements \eqref{eq:heterogeneous-so} as the unique Nash equilibrium.
\end{proposition}

\begin{proof}
The private first-order conditions are
$\mu_i-(2L+h_i)x_i-LQ_{-i}=0$, which stack as
$[H+LI+L\one\one^\top]x=\mu$. The coefficient matrix is positive definite, proving existence and uniqueness and giving \eqref{eq:heterogeneous-ne}. Aggregate welfare is
$\mu^\top x-L(\one^\top x)^2-(1/2)x^\top Hx$; strict concavity and its first-order condition give \eqref{eq:heterogeneous-so}. The additional toll contributes $-LQ_{-i}$ to allocator $i$'s derivative, so the private condition becomes $\mu_i-h_ix_i-2L\one^\top x=0$, identical to the social condition.
\end{proof}

A stand-alone $x_i^2$ penalty reduces positions but generally misses the cross-agent term in \eqref{eq:heterogeneous-so}. The empirical penalty below is a reduced-form decision rule motivated by this externality; estimating the Pigouvian toll itself requires other agents' exposures and market-impact data.

\subsection{When learning from affected returns creates a loop}

The static wedge becomes a dynamic problem when allocators learn from returns partly caused by their own earlier actions. The loop is simple: a return changes the next signal, aligned trades follow, price impact changes the next return, and the learner observes that affected return. Let unconstrained desired exposure be
\begin{equation}
 d_t=a(s+\rho r_{t-1})+br_{t-1}
 =as+(a\rho+b)r_{t-1}.
 \label{eq:desired}
\end{equation}
Here $\rho$ maps mark-to-market returns into the next signal and $b$ is a direct return response. At time $t$, define the collision-regularized allocation as the solution of
\begin{equation}
 \min_{x_1,\ldots,x_N}
 \frac12\sum_{i=1}^N(x_i-d_t)^2+\frac{c}{2}Q_t^2,
 \qquad Q_t=\sum_i x_i,\qquad c\ge0.
 \label{eq:planner}
\end{equation}
This program supplies the common-action quadratic wrapper used to isolate the feedback channel.

\begin{proposition}[Feedback stability threshold]
\label{prop:stability}
The solution of \eqref{eq:planner} is symmetric with
$x_{i,t}=d_t/(1+cN)$. Under return equation \eqref{eq:return}, deviations from the deterministic fixed point obey
\begin{equation}
 \widetilde r_t=-\frac{\lambda N(a\rho+b)}{1+cN}\widetilde r_{t-1}
 +\varepsilon_t.
 \label{eq:ar}
\end{equation}
The noiseless fixed point, and the second moment under finite-variance white noise, are stable if and only if
\begin{equation}
 \frac{|\lambda N(a\rho+b)|}{1+cN}<1.
 \label{eq:threshold}
\end{equation}
If $\kappa:=|\lambda N(a\rho+b)|>1$, stability is restored by every
$c>(\kappa-1)/N$.
\end{proposition}

\begin{proof}
The first-order condition for \eqref{eq:planner} is $x_i-d_t+cQ_t=0$.
Summing gives $Q_t=Nd_t/(1+cN)$ and therefore $x_{i,t}=d_t/(1+cN)$.
Substitution into \eqref{eq:return} and centering at the fixed point gives
\eqref{eq:ar}. The scalar AR(1) stability condition is that the absolute
coefficient be below one, which is equivalent to \eqref{eq:threshold}.
\end{proof}

The sign in \eqref{eq:ar} produces oscillatory deleveraging when $a\rho+b>0$: a positive return attracts exposure, which depresses the next return. For the stated recursion, stability is governed by the product of four measurable ingredients---impact, number of aligned allocators, signal feedback, and residual common exposure.

\subsection{When can prices reveal the common trade?}

The theory takes the common direction as known, whereas the public-market experiment must estimate a proxy from returns. The bridge from mechanism to measurement begins with the $m$-asset return vector
\begin{equation}
 R_t=u_t-\lambda vq_t,\qquad \|v\|_2=1,
 \label{eq:spike}
\end{equation}
where $u_t$ contains ordinary factors and idiosyncratic shocks, while $q_t$ is liquidation along the crowded direction $v$.

\begin{assumption}[Separated collision spike]
\label{ass:spike}
$u_t$ and $q_t$ are centered and uncorrelated. The direction $v$ is an eigenvector of $\Sigma_0=\operatorname{Cov}(u_t)$ with eigenvalue $\mu_v$. Let
$\alpha=\lambda^2\operatorname{Var}(q_t)$ and
$g=\mu_v+\alpha-\lambda_{\max}(\Sigma_0|_{v^\perp})>0$.
\end{assumption}

\begin{proposition}[Spectral certificate]
\label{prop:spectral}
Under Assumption~\ref{ass:spike}, $v$ is the leading population eigenvector of
$\Sigma_R=\operatorname{Cov}(R_t)$. If a covariance estimate satisfies
$\|\widehat\Sigma_R-\Sigma_R\|_2\le\delta<g/2$, then its leading unit eigenvector $\widehat v_1$ obeys
\begin{equation}
 \sin\angle(\widehat v_1,v)\le\frac{2\delta}{g}.
 \label{eq:dk}
\end{equation}
\end{proposition}

\begin{proof}
Uncorrelatedness gives $\Sigma_R=\Sigma_0+\alpha vv^\top$. Because $v$ is an eigenvector of $\Sigma_0$, the spike changes only its eigenvalue, from $\mu_v$ to $\mu_v+\alpha$, and the gap assumption makes it leading. Weyl's inequality implies that the two largest sample eigenvalues remain separated by at least $g-2\delta>0$, so the sample leading direction is unique up to sign. Davis--Kahan then gives $\sin\angle(\widehat v_1,v)\le 2\delta/g$ \cite{DavisKahan1970,Yu2015}.
\end{proof}

The assumptions delimit the interpretation. If flows covary with fundamental factors, cross terms appear. If $v$ is not an eigenvector of $\Sigma_0$, the population leading eigenvector mixes collision activity with ordinary risk. If the eigengap $g$ closes, the certificate weakens because the estimated direction may rotate sharply, independently of the realized loading magnitude. Consequently, the empirical loading
\begin{equation}
 c_{j,t}=|\widehat v_{1,j,t}|
 \label{eq:proxy}
\end{equation}
is a predictable \emph{collision proxy}. Identifying common AI holdings requires the filings-based alignment test in Section~\ref{sec:claims}; absolute values remove arbitrary eigenvector orientation and target the magnitude of participation in the common mode.

\section{From Any Learner to a Collision-Aware Action}

The layer sits after prediction and wraps the learner without retraining. Let $m_t\in\R^m$ be any cross-sectional score delivered by a model. A continuous relaxation makes the correction precise:
\begin{equation}
 \max_{\one^\top w=1}
 m_t^\top w-\frac{\gamma}{2}w^\top\Sigma_tw
 -\frac{\phi}{2}(c_t^\top w)^2,\qquad \phi\ge0.
 \label{eq:continuous}
\end{equation}
Define $A_\phi=\gamma\Sigma_t+\phi c_tc_t^\top$ and
\begin{equation}
 \nu_\phi=\frac{\one^\top A_\phi^{-1}m_t-1}
 {\one^\top A_\phi^{-1}\one}.
 \label{eq:nu}
\end{equation}
When $\Sigma_t\succ0$, the unique solution is
$w_\phi=A_\phi^{-1}(m_t-\nu_\phi\one)$. This expression includes the budget correction omitted by an unconstrained Sherman--Morrison argument.

\begin{proposition}[Monotone removal of collision exposure]
\label{prop:monotone}
Let $w_0$ solve \eqref{eq:continuous} at $\phi=0$. If
$c_t^\top\Sigma_t^{-1}\one=0$, then
\begin{equation}
 c_t^\top w_\phi=
 \frac{c_t^\top w_0}
 {1+(\phi/\gamma)c_t^\top\Sigma_t^{-1}c_t}.
 \label{eq:monotone}
\end{equation}
Thus $|c_t^\top w_\phi|$ is nonincreasing in $\phi$.
\end{proposition}

\begin{proof}
The orthogonality condition makes $A_\phi^{-1}\one=\gamma^{-1}\Sigma_t^{-1}\one$, so \eqref{eq:nu} equals its value at $\phi=0$. Applying Sherman--Morrison to $A_\phi^{-1}(m_t-\nu_0\one)$ and premultiplying by $c_t^\top$ yields \eqref{eq:monotone}.
\end{proof}

Equation~\eqref{eq:monotone} applies when the budget direction is orthogonal to the penalized direction in the $\Sigma_t^{-1}$ geometry. The empirical implementation uses a simpler rank approximation. At Friday $t$, compute 52-week momentum and the leading eigenvector of the trailing 104 weekly returns, standardize both across assets, and score
\begin{equation}
 z^{\mathrm{CA}}_{j,t}=z^{\mathrm{mom}}_{j,t}-\phi z^{\mathrm{PC1}}_{j,t}.
 \label{eq:score}
\end{equation}
The portfolio holds the top three assets equally over week $t+1$. In plain terms, the base model proposes attractive assets and the second term discounts assets that participate most strongly in the recent common mode. This rank rule is a model-agnostic approximation to Proposition~\ref{prop:monotone} with one discretionary coefficient.

\section{A Frozen, Prequential Experiment}

\subsection{Data, filtration, and accounting}

The frozen adjusted-close panel contains SPY, QQQ, IWM, EFA, EEM, VNQ, IEF, TLT, GLD, DBC, and SHY from January 2006. The ten non-SHY ETFs form the risky universe; SHY supplies the weekly cash proxy used in Sharpe ratios. The source CSV has SHA-256 prefix \texttt{8efbf1c8aef2969}. Prices are sampled at the last observation of each completed Friday week. Because the panel ends Thursday, July 30, 2026, the incomplete week labelled July 31 is excluded. The untouched evaluation has 760 returns from January 6, 2012 through July 24, 2026.

Every decision uses returns strictly before the held week and at least 104 prior weeks. Cross-sectional normalization is recomputed at each date, constituents stay fixed, and outcomes do not overlap. The ETF list was selected for a long common history, creating the survivorship limitation summarized in Section~\ref{sec:claims}.

Let $\mathcal F_t$ contain all completed weekly returns through Friday $t$. Momentum, covariance, eigenvectors, cross-sectional standardization, and the target $w_t$ are $\mathcal F_t$-measurable. The reported return is earned from Friday $t$ to Friday $t+1$. Adjusted closes provide a total-return convention for splits and distributions. This indexing prevents signal look-ahead; closing execution and vendor revisions remain implementation limitations.

Trading is exactly self-financing under the stated proportional-cost model. If $p_t$ is the drifted pre-trade risky fraction and $w_t$ is a fully invested target proportion, the code solves for $a_t\in(0,1]$ such that
\begin{equation}
 a_t+k\|a_tw_t-p_t\|_1=1,\qquad k=10^{-3}.
 \label{eq:self}
\end{equation}
The post-trade dollar position is $a_tw_t$; the remaining wealth pays costs. For asset return $R_{t+1}$,
\begin{equation}
 G_{t+1}=a_tw_t^\top(\one+R_{t+1}),\qquad
 p_{t+1}=\frac{a_tw_t\odot(\one+R_{t+1})}{G_{t+1}}.
 \label{eq:accounting}
\end{equation}
This avoids subtracting costs twice and measures traded notional as
$\|a_tw_t-p_t\|_1$.

\begin{table*}[t]
\caption{The 760-week design fixes the information set, accounting, diagnostics, and remaining validation targets.}
\label{tab:contract}
\centering\small
\setlength{\tabcolsep}{4.2pt}
\renewcommand{\arraystretch}{1.07}
\begin{tabular}{L{.13\textwidth}L{.25\textwidth}L{.29\textwidth}L{.22\textwidth}}
\toprule
Element & Frozen design & Implemented check & Remaining limitation\\
\midrule
Information & 52-week momentum; 104-week covariance & Inputs through Friday $t$ only; return earned over $t$ to $t+1$ & Closing execution; vendor revisions\\
Universe & Ten risky ETFs; SHY only as cash proxy & Common history, tickers, and CSV hash recorded & Ex-post universe; survivorship\\
Portfolio map & Contemporaneous $z$-scores; equal-weight top three & Stable sort; no full-sample moments & Discontinuous ranks; fixed breadth\\
Costs & 10 bp per dollar bought or sold & Equation~\eqref{eq:self} solved once; holdings drift before rebalancing & No spread, nonlinear impact, or capacity\\
Primary endpoint & Drawdown difference versus momentum & Common dates; paired blocks rebuild both wealth paths & One path; nonregular tail statistic\\
Attribution & Volatility match, PC2, 200 permutations, ten omissions & Frozen data and seeds regenerate every diagnostic & No holdings-based causal validation\\
Specification & Seven-value $\phi$ grid reported in full & JSON retains every cell, including deterioration & $\phi$ was not preregistered\\
\bottomrule
\end{tabular}
\end{table*}

\subsection{Estimands and performance definitions}

The primary empirical estimand is the difference in maximum-drawdown severity between the collision-aware and base momentum wealth paths on the frozen interval. If $W_t=\prod_{s\le t}(1+r_s)$, maximum drawdown is
\begin{equation}
 \operatorname{MDD}(r)=\min_t\left\{\frac{W_t}{\max_{u\le t}W_u}-1\right\}.
 \label{eq:mdd}
\end{equation}
We report
$\Delta_{\mathrm{DD}}=\operatorname{MDD}(r^{\mathrm{CA}})-
\operatorname{MDD}(r^{\mathrm{mom}})$, so a positive number means a less severe loss. This endpoint follows from the fire-sale mechanism's focus on common-direction stress. It was chosen after development began and is therefore exploratory; the next test freezes it prospectively.

Secondary endpoints are compound annual growth, annualized volatility, excess-return Sharpe, traded notional, and modeled cost. Compound annual growth is $W_T^{52/T}-1$. Volatility is the sample standard deviation of weekly net returns times $\sqrt{52}$. Sharpe is the weekly mean of $r_t-r_t^{\mathrm{SHY}}$ divided by its sample standard deviation, times $\sqrt{52}$. These definitions keep the compound-growth ratio separate from Sharpe and treat the 760-week record as one dependent path.

The equal-weight portfolio is a scale reference \cite{DeMiguel2009}; the second-PC and permuted-PC rules are mechanism placebos. The volatility-matched series is an ex-post attribution diagnostic because its scale uses full-period volatility. Table~\ref{tab:contract} fixes these roles and separates investable paths from falsification checks.

The return and drawdown estimands incorporate the exact cost recursion \eqref{eq:self}. Traded notional is bought-plus-sold notional divided by pre-trade wealth. Thus a complete switch between two fully invested names has notional close to two and costs close to 20 basis points, reconciling the code, tables, and prose.

\subsection{Why the wrapper is evaluated prequentially}

Rolling principal components are sign-indeterminate and regime-dependent. At each decision we orient neither eigenvector; equation \eqref{eq:proxy} uses absolute loadings. The cross-sectional standardization at $t$ uses only the ten contemporaneous loadings, so it cannot import future scale. Eigenvalue magnitude cancels under this standardization; the implementation therefore uses $|\widehat v_{1,j,t}|$, while the leading eigenvalue share is retained only as a diagnostic. Ties are broken by a stable asset order. These conventions make the rank path deterministic.

The experiment fixes momentum as the base score and changes only its common-mode alignment. This isolates the wrapper from representation learning, hyperparameter optimization, and training randomness. A neural or agentic allocator can occupy the same proposal interface; validating AI-specific crowding requires the model-lineage and holdings data specified in Section~\ref{sec:claims}.

There are two sources of adaptation. The 52-week score changes the proposed direction, and the 104-week covariance changes the collision proxy. Both are slow relative to the weekly holding interval. Each estimate precedes its evaluated outcome, which makes the construction prequential. The paired bootstrap addresses short-range dependence shared by both paths while leaving stationarity and randomized-intervention assumptions outside the design.

\begin{table*}[t]
\caption{One prequential update. Every object used in Steps 1--6 is observed before the held return in Step 7.}
\label{tab:algorithm}
\centering\small
\setlength{\tabcolsep}{5pt}
\renewcommand{\arraystretch}{1.07}
\begin{tabular}{clp{.68\textwidth}}
\toprule
Step & Object & Operation\\
\midrule
1 & Information & At completed Friday $t$, retain weekly returns only through $t$ and the drifted pre-trade fractions $p_t$.\\
2 & Base score & Compound each asset's last 52 weekly returns and cross-sectionally standardize the resulting momentum vector.\\
3 & Collision mode & Estimate the sample covariance from the last 104 weeks; take absolute loadings of its leading unit eigenvector.\\
4 & Comparable scale & Cross-sectionally standardize those ten absolute loadings using only date-$t$ values.\\
5 & Wrapped decision & Compute \eqref{eq:score}, apply the stable top-three ranking, and assign one third to each selected ETF.\\
6 & Self-financing trade & Solve \eqref{eq:self} by monotone bisection; record bought-plus-sold notional and the associated cost once.\\
7 & Held outcome & Apply only the next Friday-to-Friday asset-return vector to the financed positions and update wealth.\\
8 & State update & Divide marked positions by new wealth to obtain $p_{t+1}$; repeat without reusing the held return in the earlier score.\\
\bottomrule
\end{tabular}
\end{table*}

For $m$ assets, the dominant numerical operation is the eigendecomposition of an $m\times m$ covariance matrix, $O(m^3)$ with a dense solver; score construction and trading are linear after sorting. The experiment has $m=10$, so numerical error is negligible relative to modeling error. The self-financing equation is continuous and strictly increasing in $a$ apart from harmless absolute-value kinks, and its left side is at most one at $a=0$ and at least one at $a=1$ when a trade occurs. Bisection therefore returns the unique maximal financed scale to machine tolerance.

\begin{table}[t]
\caption{The wrapper's observed trade-off is lower drawdown and trading with essentially unchanged Sharpe ($T=760$ weeks; 10 bp per dollar traded). CAGR, volatility, and MDD are percentages; Trade/wk is mean bought-plus-sold notional relative to pre-trade wealth.}
\label{tab:main}
\centering\small
\setlength{\tabcolsep}{2.5pt}
\renewcommand{\arraystretch}{1.07}
\begin{tabular}{@{}lrrrrr@{}}
\toprule
Method & CAGR & Vol. & Sharpe & MDD & Trade/wk\\
\midrule
Equal weight & 8.43 & 10.87 & 0.686 & $-22.73$ & 0.013\\
Momentum & 9.18 & 12.62 & 0.664 & $-23.60$ & 0.241\\
Collision-aware & 7.98 & 10.68 & 0.667 & $-14.14$ & 0.193\\
Mom., vol.-matched$^\dagger$ & 7.95 & 10.56 & 0.664 & $-19.89$ & 0.201\\
PC2 placebo & 11.08 & 13.60 & 0.746 & $-23.37$ & 0.189\\
\bottomrule
\end{tabular}
\parbox{.98\columnwidth}{\vspace{1mm}\small\emph{Note.} Sharpe uses weekly returns in excess of SHY. Collision-aware uses $\phi=0.75$. $^\dagger$Ex-post attribution diagnostic, not an investable strategy: momentum excess returns and turnover are multiplied by 0.834.}
\end{table}

\subsection{Primary estimates and uncertainty}

Table~\ref{tab:main} shows the observed trade-off. At $\phi=0.75$, collision awareness gives up 1.20 annual return points and 1.94 volatility points relative to momentum. Maximum drawdown improves by 9.47 points, mean weekly traded notional falls 20.0\%, and annual modeled cost falls from 1.25\% to 1.00\% of wealth. Excess-return Sharpe changes by only 0.004. These are path statistics, not independent observations.

We use the paired stationary bootstrap \cite{PolitisRomano1994}: the same circular blocks of both strategies and SHY are sampled, with expected block length 13 weeks and 2,000 deterministic-seed replications. The wealth path, maximum drawdown, and Sharpe are recomputed inside each replication. This preserves contemporaneous pairing and short-range dependence better than an iid bootstrap, but it cannot create additional crises and is not exact finite-sample inference for maximum drawdown.

\begin{table}[t]
\caption{The drawdown-gain interval includes zero; placebo and omission summaries remain descriptive ($T=760$ weeks).}
\label{tab:robust}
\centering\small
\setlength{\tabcolsep}{3.0pt}
\renewcommand{\arraystretch}{1.07}
\begin{tabular}{p{.47\columnwidth}rr}
\toprule
Diagnostic & Point & Reported range\\
\midrule
DD gain, paired bootstrap (pp) & 9.47 & [$-3.25$, 18.38]\\
$\Pr^*(\mathrm{DD\ gain}>0)$ & 0.895 & --\\
Sharpe gain, paired bootstrap & 0.004 & [$-0.324$, 0.329]\\
Permuted-PC MDD severity (\%) & 24.91 & [19.32, 33.44]\\
Leave-one-out DD gain (pp) & 5.38 & [$-0.80$, 9.28]$^\dagger$\\
\bottomrule
\end{tabular}
\parbox{.98\columnwidth}{\vspace{1mm}\small\emph{Note.} DD gain is positive when the collision-aware loss is less severe. The paired rows use 2,000 stationary-bootstrap draws and 95\% percentiles. Permuted PC uses 200 runs and their 2.5--97.5 percentiles. $^\dagger$Minimum--maximum over ten omitted-asset universes; nine gains are positive.}
\end{table}

The drawdown interval includes zero and the bootstrap probability is 0.895, so the crash-reduction estimate remains statistically unresolved. Its 9.47-point estimate is economically large and retains its sign in several attribution checks, while the Sharpe interval centers near zero. A later locked period is required to narrow uncertainty beyond this single market history.

\begin{figure*}[t]
\centering
\includegraphics[width=.94\textwidth]{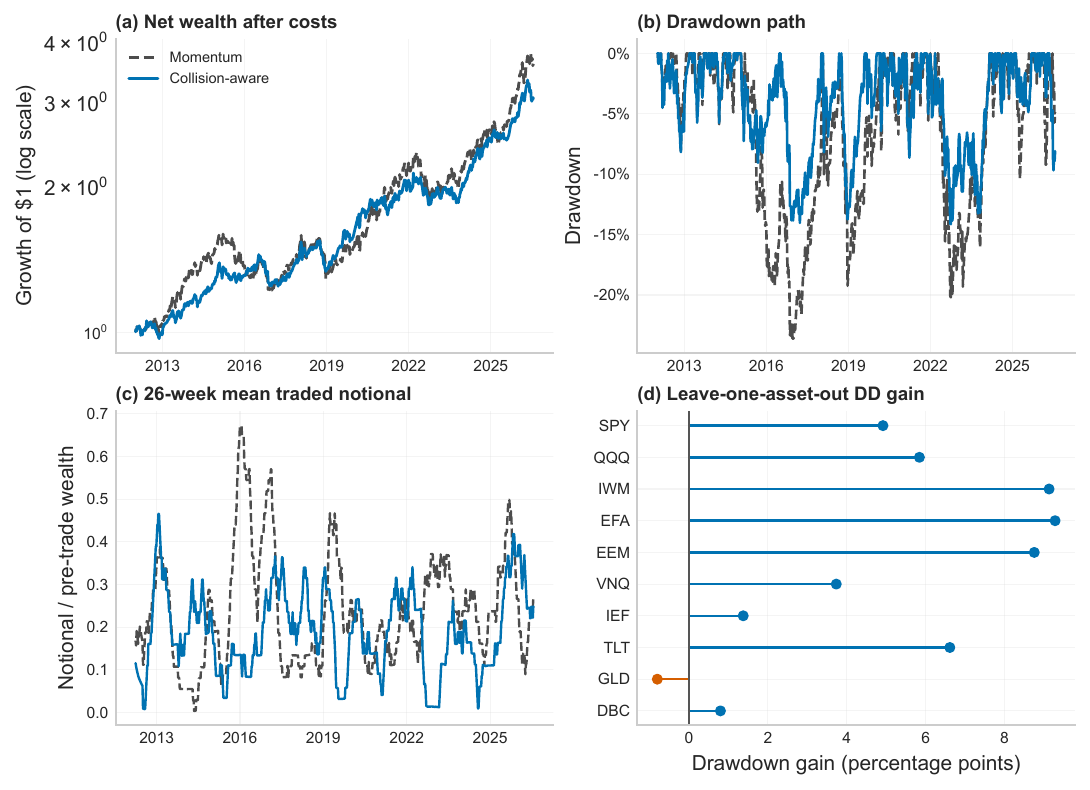}
\caption{The wrapper reduces the deepest realized loss and traded notional, while the asset-omission diagnostic retains the drawdown-gain sign in nine of ten universes. The $T=760$ wealth and drawdown paths include 10 bp proportional costs; momentum is dashed gray and collision-aware is solid blue. Panel (c) uses a 26-week mean only for display. Panel (d) reports leave-one-asset-out drawdown gains in percentage points; the vertical line marks zero and GLD is the exception.}
\label{fig:audit}
\Description{Four panels compare a dashed gray momentum path with a solid blue collision-aware path from 2012 to 2026. The wrapper has lower terminal wealth but a shallower deepest drawdown and lower smoothed traded notional. The final panel shows ten leave-one-asset-out drawdown gains: nine are positive, while omission of GLD is slightly negative.}
\end{figure*}

\subsection{Falsification and sensitivity}

Four diagnostics test whether generic de-risking can reproduce the observed path.
First, ex-post volatility matching scales momentum excess returns by 0.834. Its drawdown is 19.89\%, still 5.75 points worse than collision-aware. This diagnostic uses full-test volatility and therefore is not a tradable competitor; it isolates scale from path composition.

Second, penalizing the second covariance eigenvector produces higher return and Sharpe but a 23.37\% drawdown. Thus a penalty on an arbitrary low-rank direction does not reproduce the primary tail statistic. Third, in 200 placebo runs, PC1 loadings are independently permuted across assets at each date before ranking. Their median drawdown is 24.91\%, with 95\% simulation range 19.32--33.44\%. The primary reduction is therefore not reproduced by an arbitrary noisy cross-sectional score. The placebo distribution is conditional on the observed history and randomized ranks; it is not a sampling confidence interval.

Fourth, the covariance, rankings, and backtest are recomputed after omitting each risky ETF. Collision-aware drawdown improves in nine of ten universes. The median improvement is 5.38 points, the range is $-0.80$ to 9.28 points, and omission of GLD is the exception; Figure~\ref{fig:audit}(d) displays every omission. The magnitude therefore depends on asset composition, but no single bond or equity ETF alone creates the reported sign.

Table~\ref{tab:phi} displays the complete coefficient grid at
$\phi=0$, 0.25, 0.50, 0.75, 1, 1.25, and 1.5. Drawdown severity falls from 23.6\% at zero to 14.1\% at .75, then rises to 16.8\% at 1 and 20.3\% at 1.25. Sharpe peaks at .50, while .75 is the drawdown-oriented coefficient. Because $\phi$ was selected during development, the grid is descriptive; the next experiment freezes $\phi=.75$, the universe, and endpoints before new observations arrive.

\begin{table*}[t]
\caption{The complete penalty grid exposes a nonmonotone risk--return--trading frontier ($T=760$ weeks; 10 bp per dollar traded). CAGR, volatility, and MDD are percentages; Trade/wk is mean bought-plus-sold notional relative to pre-trade wealth.}
\label{tab:phi}
\centering\small
\renewcommand{\arraystretch}{1.07}
\begin{tabular*}{\textwidth}{@{\extracolsep{\fill}}rrrrrr@{}}
\toprule
$\phi$ & CAGR & Volatility & Excess-return Sharpe & MDD & Trade/wk \\
\midrule
0.00 & 9.18 & 12.62 & 0.664 & $-23.60$ & 0.241 \\
0.25 & 8.47 & 11.81 & 0.646 & $-19.30$ & 0.205 \\
0.50 & 9.27 & 11.50 & 0.730 & $-16.39$ & 0.175 \\
0.75 & 7.98 & 10.68 & 0.667 & $-14.14$ & 0.193 \\
1.00 & 6.28 & 9.74 & 0.564 & $-16.78$ & 0.171 \\
1.25 & 4.20 & 9.53 & 0.362 & $-20.32$ & 0.150 \\
1.50 & 3.35 & 9.32 & 0.276 & $-18.70$ & 0.104 \\
\bottomrule
\end{tabular*}
\end{table*}

\section{Claim Boundary and Decisive Next Test}
\label{sec:claims}

The empirical mechanism aligns with the theory along three measured dimensions: the wrapper reduces exposure to the most pervasive recent direction, lowers trade volume, and changes the stress path more than volatility-only scaling. Structural attribution requires three additional objects---time-stamped model lineage, institutional holdings overlap, and signed impact---because macro shocks, duration exposure, and changing equity--bond correlation can also generate a leading principal component.

Direct validation requires holdings, signed flow, and a later locked evaluation. Table~\ref{tab:identification} separates these objects because success on one does not identify the others.

\begin{table*}[t]
\caption{Each structural object has a distinct observable, validation data set, and rejection criterion.}
\label{tab:identification}
\centering\small
\renewcommand{\arraystretch}{1.07}
\setlength{\tabcolsep}{4.0pt}
\begin{tabular}{L{.15\textwidth}L{.22\textwidth}L{.29\textwidth}L{.26\textwidth}}
\toprule
Structural claim & Current observable & Required validation data & Evidence against the mechanism \\
\midrule
Collision direction $v$ & Rolling leading-covariance loading magnitude & Filings available at each rebalance; a pre-specified holdings-overlap direction & No out-of-sample loading alignment beyond factor and permutation controls \\
Impact $\lambda$ or $\Lambda$ & Not identified; return covariance is only a proxy & Signed flow or order imbalance, liquidity, and subsequent price response & Nonpositive impact, or no larger response along the proposed direction \\
Feedback gain & Return dynamics do not separate impact, allocator count, and updating gain & Allocator capital, update rules, and impact estimates measured before the outcome & Stable loop without attenuation when common-mode exposure is reduced \\
Causal AI crowding & No direct observable in the present artifact & Time-stamped model use or lineage joined to holdings or orders & Comparable overlap among non-AI controls, or disappearance after fundamental-factor controls \\
\bottomrule
\end{tabular}
\end{table*}

\subsection{Deployment and identification ladder}

Each deployment stage has its own artifact and gate.
\emph{Stage 1: auditable action control.}
Record the base score, eigenvector and eigengap, $\phi$, pre/post loading, holdings, financed scale, traded notional, cost, and next return.
This ledger reconstructs every correction and supports live limits on loading, turnover, and numerical residuals.

\emph{Stage 2: proxy validation.}
Freeze the price-based direction, then use only N-PORT or 13F filings public at each rebalance to construct a holdings-overlap direction.
Pre-specify the sign-invariant angle between the two directions and compare it with fundamental-factor directions, PC2, and permuted loadings.
Report filing coverage, publication lags, the alignment path, and results stratified by eigengap.
Forecasting later overlap beyond these controls validates the proxy bridge; failure supplies the rejection criterion in Table~\ref{tab:identification}.

\emph{Stage 3: impact validation.}
Join the frozen direction to signed fund flow or order imbalance, contemporaneous liquidity, and subsequent return.
Estimate whether flow predicts larger adverse price response along it than along matched orthogonal directions.
A nonpositive differential rejects the impact channel; time-stamped model lineage joined to holdings or orders supports AI-specific attribution.

\emph{Stage 4: prospective decision value.}
Lock the universe, 52/104-week windows, breadth, $\phi=0.75$, cost rule, and drawdown endpoint before a new period.
Report every intervention, the wealth path, and the same volatility, eigenmode, permutation, omission, and bootstrap diagnostics.
The four stages earn progressively stronger conclusions: executable risk control, validated overlap proxy, measured impact mechanism, and AI-specific attribution.

The return panel introduces further limitations. Adjusted closes omit bid--ask spreads, market impact, and execution at a volume-weighted price; linear costs are exact within the declared model. The top-three rule is discontinuous, which permits nonmonotone turnover in $\phi$. The universe is small, long-only, unlevered, and selected with hindsight. SHY is a traded Treasury ETF rather than a risk-free account. Weekly sampling misses intraperiod drawdown and liquidation. Extension to individual equities or faster trading requires a new execution model and evaluation.

The game is stylized: the baseline and feedback models are symmetric, impact is linear, the common direction is exogenous, and strategic disclosure, entry, and leverage constraints are absent. Proposition~\ref{prop:heterogeneous} relaxes static symmetry while retaining linear impact and unconstrained scalar actions. Proposition~\ref{prop:game} supplies a local externality benchmark; calibrated welfare requires estimated impact and exposure. Proposition~\ref{prop:stability} applies to recursion \eqref{eq:desired}, and Proposition~\ref{prop:spectral} relies on orthogonality and an eigengap that may weaken during regime changes.

The contribution is a model-independent decision layer, a closed-form economic rationale, and an audit of its proposed mechanism. The primary path is deterministic; the artifact freezes the CSV hash, completed-week cutoff, filtration, cost equation, performance definitions, bootstrap seed, placebo seed, and every coefficient in the displayed grid. These records reconstruct every reported claim from the frozen input.

\subsection{Testable predictions and more than one direction}

The results yield four pre-specified tests for a later, richer data set. First, the aggregate private--social exposure gap approaches the nonzero limit in Corollary~\ref{cor:welfare} as the number of aligned allocators grows, although each allocator's gap shrinks. Second, feedback risk scales with the product of impact and return-responsive updating in \eqref{eq:threshold}, which the holdings-and-flow design must estimate. Third, a collision penalty can reduce stress loss while leaving unconditional Sharpe essentially unchanged, as in Table~\ref{tab:main}. Fourth, a closing eigengap weakens the spectral certificate and permits greater rotation of the estimated direction independently of the realized proxy magnitude.

Several shared directions can be handled by replacing scalar impact $L$ with a positive-semidefinite impact matrix $\Lambda$ and applying one coefficient per retained direction. This study fixes one direction to avoid rank selection on the test path; the PC2 placebo also shows that extra modes can miss the observed drawdown. A future design should freeze the rank using past data and validate every retained direction against holdings overlap.

Collision control also differs from covariance shrinkage. Shrinkage estimates exogenous risk; collision control treats part of risk as induced by aggregate action. The decisive test asks whether the penalized direction forecasts later portfolio overlap, flow sensitivity, or price pressure after controlling for fundamental factors.

\section{Conclusion}
\enlargethispage{\baselineskip}

The market sees actions, not model names. In the collision game, private allocators overuse a shared direction because each ignores part of the impact imposed on others; an overlap toll implements the social choice. When allocation reacts to affected returns, the same common component determines a precise stability threshold. Under a separated-spike model, a leading eigenportfolio supplies a measurable reduced-form proxy for that component.

In the frozen ETF test, penalizing the leading-mode loading reduces the deepest realized drawdown by 9.47 percentage points and weekly traded notional by 20.0\%, with essentially unchanged Sharpe. Volatility matching and eigenmode, permutation, and omission diagnostics reject several simple explanations, while the zero-crossing block-bootstrap interval records the remaining sampling uncertainty. The result delivers a reproducible collision-aware layer and a decisive next experiment: freeze the rule, validate its direction against holdings overlap, and test signed price impact on later flows.
The same audit artifacts make common-mode exposure measurable for model governance, capacity limits, and cross-manager stress tests.

\end{document}